\documentclass[10pt, conference]{IEEEtran}
\IEEEoverridecommandlockouts

\usepackage{cite}
\usepackage{amsmath,amssymb,amsfonts}
\usepackage{algorithmic}
\usepackage{graphicx}
\usepackage{textcomp}
\usepackage{xcolor}
\usepackage{authblk}

\usepackage{hyperref}

\usepackage{tikz,tkz-tab}
\usepackage{physics}
\usepackage{caption}
\usepackage{subcaption}

\def\BibTeX{{\rm B\kern-.05em{\sc i\kern-.025em b}\kern-.08em
    T\kern-.1667em\lower.7ex\hbox{E}\kern-.125emX}}

\usepackage{mathbbol} 

\newcommand{\dom}{\Omega}
\newcommand{\normal}{\boldsymbol{\nu}}
\newcommand{\mesh}{\mathcal{T}_h}
\newcommand{\sumK}{\sum_{K\in\mesh}}

\newcommand{\jumpT}[1]{\left[ #1 \right]_T}
\newcommand{\jumpN}[1]{\left[ #1 \right]_N}

\newcommand{\Ltwospace}[1]{L^2(#1)}
\newcommand{\Ltwodom}{\Ltwospace{\dom}}
\newcommand{\Ltwodomvec}{\left(\Ltwospace{\dom}\right)^d}

\newcommand{\normLtwo}[2]{|| #1 ||_{\Ltwospace{#2}}}
\newcommand{\normLtwovec}[2]{|| #1 ||_{\left(\Ltwospace{#2}\right)^d}}

\newcommand{\normLtwoK}[1]{\normLtwo{#1}{K}}
\newcommand{\normLtwoKvec}[1]{\normLtwovec{#1}{K}}
\newcommand{\normLtwodomvec}[1]{\normLtwovec{#1}{\dom}}

\newcommand{\Hrot}{\mathbf{H}(\text{curl})}
\newcommand{\Hrotspace}[1]{\mathbf{H}(\text{curl},#1)}
\newcommand{\Hrotdom}{\Hrotspace{\dom}}
\newcommand{\Hrotzerodom}{\mathbf{H}_0(\text{curl},\dom)}

\renewcommand{\grad}{\boldsymbol{\nabla}}
\renewcommand{\div}{\nabla \cdot}
\newcommand{\rot}{\boldsymbol{\nabla} \times }

\newcommand{\froemint}{\Gamma}

\newcommand{\E}{\mathbf{E}}
\newcommand{\Eh}{\E_h}
\newcommand{\Ehvec}{\underline{E}_h}
\newcommand{\Fbf}{\mathbf{F}}
\newcommand{\Fvec}{\underline{F}}

\newcommand{\ubf}{\mathbf{u}}
\newcommand{\vbf}{\mathbf{v}}
\newcommand{\xbf}{\mathbf{x}}

\newcommand{\bigO}{\mathcal{O}}

\newtheorem{Th}{Theorem}[section]
\newtheorem{Def}{Definition}[section]

\begin{document}

\title{Adaptive mesh refinement quantum algorithm for Maxwell's equations}

\author[1,2]{Elise Fressart}
\author[1]{Michel Nowak}
\author[2]{Nicole Spillane}

\affil[1]{cortAIx Labs, Thales Research and Technology, 91120 Palaiseau, France}
\affil[2]{CMAP, CNRS, École polytechnique, Institut Polytechnique de Paris, 91120 Palaiseau, France.}

\maketitle
\let\thefootnote\relax\footnotetext{elise.fressart@thalesgroup.com}
\let\thefootnote\relax\footnotetext{michel.nowak@thalesgroup.com}
\let\thefootnote\relax\footnotetext{nicole.spillane@cmap.polytechnique.fr}

\begin{abstract}
Algorithms that promise to leverage resources of quantum computers efficiently to accelerate the finite element method have emerged.
However, the finite element method is usually incorporated into a high-level numerical scheme which allows the adaptive refinement of the mesh on which the solution is approximated.
In this work, we propose to extend adaptive mesh refinement to the quantum formalism, and apply our method to the resolution of Maxwell's equations.
An important step in this procedure is the computation of error estimators, which guide the refinement.
By using block-encoding, we propose a way to compute these estimators with quantum circuits.
We present first numerical experiments on a 2D geometry.

\end{abstract}


\section{Introduction}
Partial differential equations (PDEs) appear in various scientific fields such as fluid dynamics, mechanics, electromagnetism. The finite element method (FEM) \cite{zbMATH03596197,zbMATH05223061} is a conventional approach to numerically solve PDEs.
After discretizing the domain, the FEM builds a large linear system whose unknowns represent the degrees of freedom of the solution.
Solving such systems can be very resource demanding on a classical computer even if they are sparse. 

The Harrow-Hassidim-Lloyd (HHL) algorithm \cite{PhysRevLett.103.150502} is the first quantum algorithm to solve the quantum linear system problem (QLSP). Given a quantum state $\ket{b}$ and a matrix $A$, the algorithm outputs, up to some precision, a quantum state $\ket{x}$ proportional to the solution $x$ of the linear system $A x = b$. Under suitable assumptions, the total runtime of HHL  depends logarithmically on the size $N$ of the problem. This is an exponential speedup compared to a classical iterative solver like the conjugate gradient method \cite{zbMATH01953444}.
More recently, variable-time amplitude amplification \cite{ambainis:LIPIcs.STACS.2012.636}, linear combination of unitaries (LCU) \cite{doi:10.1137/16M1087072}, and block-encoding associated with quantum singular value transformation (QSVT) \cite{Gilyen_2019} propose to improve the complexity.
Recent research has focused on algorithms inspired by, or based on, adiabatic quantum computing \cite{PhysRevLett.122.060504, 10.1145/3498331, Lin2020optimalpolynomial, PRXQuantum.3.040303}. An optimal complexity, with respect to the precision $\epsilon$ and the condition number $\kappa$, of the matrix $A$ is reached in \cite{PRXQuantum.3.040303}.
The integration of a QLSP solver into the FEM has been proposed by~\cite{montanaro2016quantum} where a second order elliptic PDE is used to illustrate the new numerical scheme.
An application to electromagnetic simulations is proposed in \cite{PhysRevLett.110.250504}.

Near-term variational algorithms to solve the QLSP have also received growing attention. The work in \cite{e25040580} applies the variational quantum linear solver \cite{BravoPrieto2023variationalquantum} to the FEM. 

However, the integration of a QLSP solver into the FEM might be lacking one important consideration: the mesh on which the solution is discretized is often chosen arbitrarily.
In order to design a mesh that captures all the variations of the solution properly, we rely on classical adaptive schemes \cite{zbMATH05707381}.
They allow to reduce the computational cost by optimizing the placement of the degrees of freedom in order to decrease the error.
The adaptive schemes refine iteratively the cells of the mesh where the error is large.
This can be done by applying a posteriori error estimation \cite{https://doi.org/10.1002/nme.1620121010, zbMATH00911319}, which aims at approximating the error distribution from the lastly computed solution.

In this work, we focus on Maxwell's propagation equations. 
We suppose that we have access to a quantum linear system solver that efficiently outputs the degrees of freedom of the  FEM in a quantum register.
From there, we reformulate the local error estimators and propose quantum circuits to compute them directly.
We then iterate on the refined meshes and show that global error is indeed reduced compared to a uniform refinement strategy.

The rest of the article is organized as follows. Section \ref{sec:classical_approch} presents the classical adaptive mesh refinement algorithm relying on residual-based a posteriori error estimation for Maxwell's equation. In Section \ref{sec:hybrid_amr}, we propose a hybrid algorithm for adaptive mesh refinement. Section \ref{sec:numerical_simulations} focuses on numerical simulations. In Section \ref{sec:conclusion}, we summarize our findings and offer up some perspectives for future work.

\section{Classical approach}
\label{sec:classical_approch}
\subsection{Maxwell's equations}
We start by recalling Maxwell's equations in a linear, non-dispersive, isotropic medium with time-independent permittivity $\epsilon$, permeability $\mu$ and conductivity $\sigma$ \cite{zbMATH01868846}. These equations introduce the electric field $\E(\xbf,t)$ and the magnetic field $\mathbf{B}(\xbf,t)$ for all $\xbf \in \mathbb{R}^3$ and $t \in \mathbb{R}$. The first equation relates the divergence of the electric field $\mathbf{E}$ with the charge density $\rho$ by
\begin{equation}
\div \left( \epsilon \E \right)=\rho \text.
\label{eq:gauss}
\end{equation}
The second equation states that the divergence of the magnetic field is null
\begin{equation}
\div \mathbf{B}=0\text.
\label{eq:gauss_magnetism}
\end{equation}
The third equation relates the curl of the electric field to the time derivative of the magnetic field as follows
\begin{equation}
\frac{\partial\mathbf{B}}{\partial t} + \rot \E =0 \text.
\label{eq:faraday}
\end{equation}
Finally, the last equation is between the curl of the magnetic field, the current density $\mathbf{J}$, which under Ohm's law with applied current density $\mathbf{J}_e$ equals $\mathbf{J}_e + \sigma \mathbf{E}$, and the time derivative of the electric field
\begin{equation}
\epsilon \frac{\partial\E}{\partial t} - \rot \left( \mu^{-1} \mathbf{B} \right)  = - \mathbf{J}_e - \sigma \E \text.
\label{eq:ampere_maxwell}
\end{equation}
Taking the time derivative of $\eqref{eq:ampere_maxwell}$ and switching the order of the spatial and time derivatives leads to
\begin{equation}
 \epsilon \frac{\partial^2 \E}{\partial^2 t} - \rot  \left( \mu^{-1} \frac{\partial \mathbf{B}}{\partial t} \right) = - \frac{\partial \mathbf{J_e}}{\partial t} - \sigma \frac{\partial \E}{\partial t} \text.
\end{equation}
Using \eqref{eq:faraday}, we obtain:
\begin{equation}
 \epsilon \frac{\partial^2 \E}{\partial^2 t} + \sigma \frac{\partial \E}{\partial t} + \rot  \left( \mu^{-1} \rot \E \right) = - \frac{\partial \mathbf{J_e}}{\partial t} \text.
\end{equation}
In the time-harmonic setting, the time dependence is periodic with pulsation $\omega$ so the fields, the charge density and the current density can be rewritten as $f(x,t) = f(x) \exp(-i \omega t)$ (with generic $f$ denoting any of these variables).
A time derivative multiplies by $-i \omega$. By replacing $\E$ with $\epsilon_0^{1/2} \E$ and multiplying by $\epsilon_0^{1/2} \mu_0$, we obtain the final propagation equation:
\begin{equation}
\rot( \mu_r^{-1} \rot \E)- k^2 \epsilon_r \E = i k \mu_0^{1/2} \mathbf{J_e}
\end{equation}
where $k=\omega \sqrt{\epsilon_0 \mu_0}$ is the wave number, $\epsilon_r = \frac{1}{\epsilon_0} \left(\epsilon + \frac{i \sigma}{\omega} \right)$ and $\mu_r = \frac{\mu}{\mu_0}$ are normalized by $\epsilon_0$ and $\mu_0$ (respectively the permittivity and permeability of free space).

\subsection{Propagation in a perfect conductor cavity}
\begin{figure}[htbp]
    \centering
    \begin{tikzpicture}[scale=1]
    \coordinate (E)  at(1.25,1.25) {};
    \coordinate (F)  at(2.5,1.25) {};
    \coordinate (G)  at(2.5,2.5) {};
    \coordinate (H)  at(0,2.5) {};
    \coordinate (I)  at(0,0) {};
    \coordinate (J)  at(1.25,0) {};
    
    \draw[black, fill=gray!25!white] (E) -- (F) -- (G) -- (H) -- (I) -- (J) -- cycle;

    \node[draw=none, text=black] (dom) at(1.875,1.875) {$\dom$ };
    \node[draw=none, text=black] (gam) at(-0.15,1.875) {$\froemint$ };
    \coordinate (Y) at(1.25,0.625) {};
    \coordinate (Z) at(2.25,0.625) {};
    \draw[->,line width=0.25mm] (Y) -- (Z) node[midway,above,outer sep=-10pt] {$\normal$} ;
    \end{tikzpicture}
    \caption{Example of a two-dimensional domain $\Omega$ with boundary $\Gamma$ and outer unit normal vector $\normal$.}
    \label{fig:domain_illustration}
\end{figure}
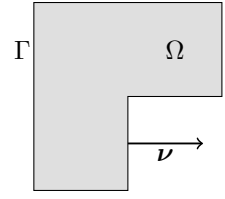

In a bounded polyhedral domain $\dom \subset \mathbb{R}^d$ ($d=2,3$) with boundary $\Gamma$ and outward unit normal $\normal$, we seek the electric field $\E$ satisfying the time-harmonic Maxwell's equation and subject to the perfect electric conductor boundary condition:
\begin{align}
    \rot (\mu_r^{-1} \rot \E)- k^2 \epsilon_r \E & = \Fbf \quad \text{in } \dom \label{eq:pbcavite1} \\
    \normal \times \E & = 0 \quad \text{on }  \froemint
    \label{eq:pbcavite}
\end{align}
where $\Fbf$ is the source term, $k$ the wave number and $\epsilon_r$ and $\mu_r$ are the relative permittivity and the relative permeability respectively. We assume that $\div F = 0$. A possible domain is represented in Fig \ref{fig:domain_illustration}.
The solution space of the problem is 
\begin{align*}
\Hrotzerodom = &\left\{ \ubf \in \Ltwodomvec ,~ \rot \ubf  \in \Ltwodomvec ~|~  \right. \\
& \left. \ubf \times \normal = 0 \text{ on } \partial \dom \right\} 
\end{align*}
equipped with the norm 
\begin{equation*}
|| \vbf ||_{\Hrotdom} = \left( \normLtwodomvec{\vbf}^2 + \normLtwodomvec{\rot \vbf}^2 \right)^{1/2},
\end{equation*}
where in turn $\normLtwodomvec{\vbf} = \left( \sum_{i=1}^d \int_\dom |\vbf_i|^2 \right)^{1/2}$.

The variational formulation is obtained by multiplying \eqref{eq:pbcavite1} by a test function in $\Hrotzerodom$ and using integration by parts. The boundary term vanishes because of the boundary condition satisfied by the test function. The variational formulation consists in finding $\E \in \Hrotzerodom$ such that 
\begin{equation}
    \int_\dom \mu_r^{-1} \rot \E \cdot \rot \overline{\vbf} - \int_\dom k^2 \epsilon_r \E \cdot \overline{\vbf} = \int_\dom \Fbf \cdot \overline{\vbf},
    \label{eq:pbcavitevar}
 \end{equation}
for all $\vbf \in \Hrotzerodom$. We assume that $k$ is not a resonant wave number so that \eqref{eq:pbcavitevar} has a unique solution for any $\Fbf \in \left(\Ltwodom\right)^d$ \cite[Corollary 4.19]{zbMATH01868846}. Let $\E$ denote this solution. \\

\subsection{The Finite Element Method}
The Finite Element Method (FEM)~\cite{zbMATH01868846, zbMATH06295908} is a standard approach to approximate the solution to \eqref{eq:pbcavitevar}. It consists in the following steps.

\subsubsection{Meshing the domain}
\begin{figure}[htbp]
    \centering
    \includegraphics[scale=0.07]{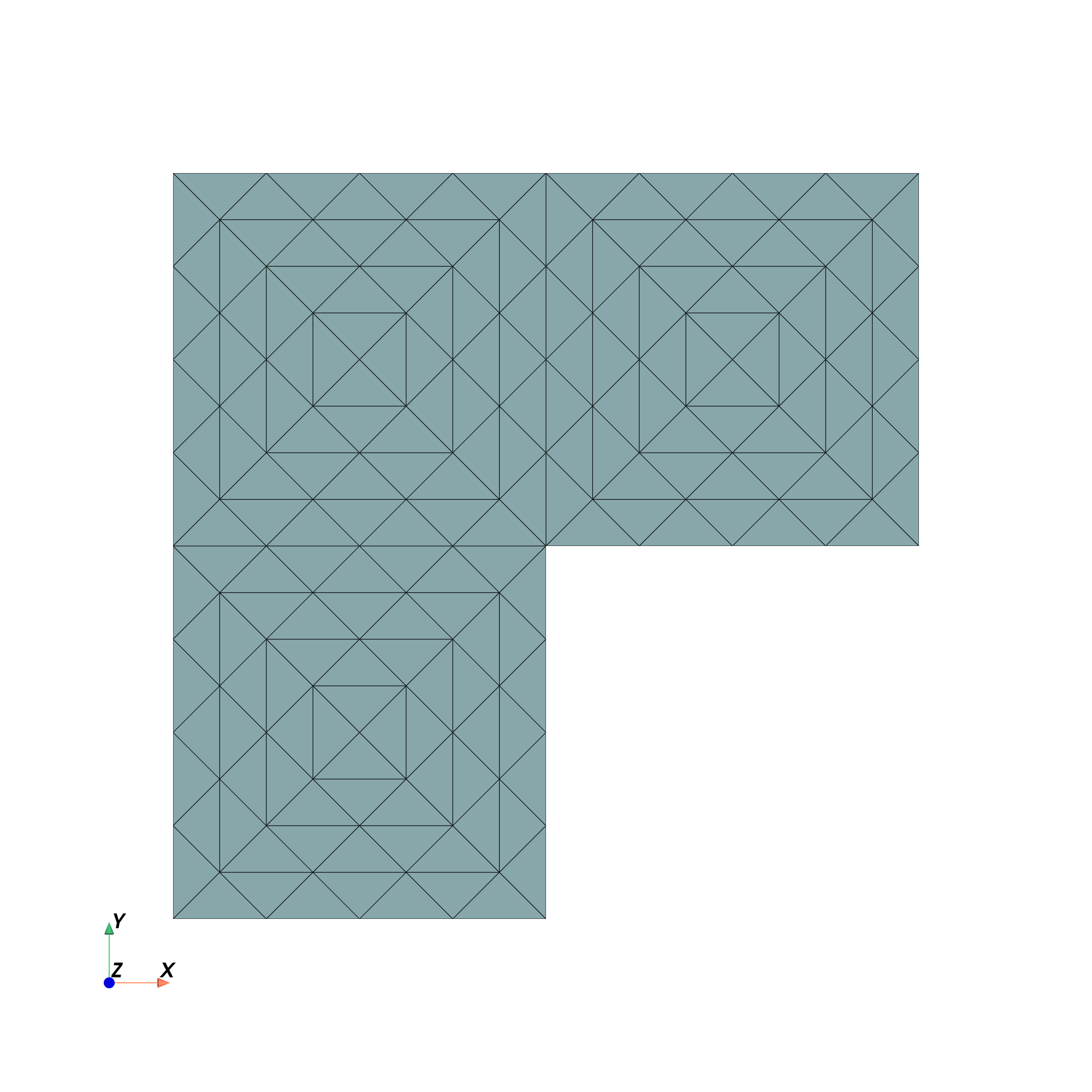}
    \caption{Illustration of a meshed domain}
    \label{fig:meshed_domain_illustration}
\end{figure}

The domain is discretized with a set of elements (e.g triangles, tetrahedra) forming a mesh $\mesh$ as in Fig.\ref{fig:meshed_domain_illustration}. 
\subsubsection{Choosing the basis functions}
\begin{figure}[h]
\begin{center}
\begin{tikzpicture}[,
squarednode/.style={rectangle, draw=black!100, very thick, minimum size=0.5cm},
]

\node[line width=1pt] (P1)  at(0,0) {};
\node[line width=1pt] (P2)  at(2,0) {};
\node[line width=1pt] (P3)  at(0,2) {};

\draw [line width=0.2mm] (0,0) -- (2,0);
\draw [line width=0.2mm] (2,0) -- (0,2);
\draw [line width=0.2mm] (0,2) -- (0,0);
\draw [blue, line width=0.4mm,-stealth](0.5,0) -- (1.5,0);
\draw [blue, line width=0.4mm,-stealth](0,1.5) -- (0,0.5);
\draw [blue, line width=0.4mm,-stealth](1.324,0.676) -- (0.676,1.324);
\end{tikzpicture}
\end{center}
\caption[Degrees of freedom for Nédélec finite element of order 1]{Degrees of freedom for Nédélec finite elements of order 1. An arrow collinear to an edge indicates that the degree of freedom is the value of the tangential component integrated over this edge.}
\label{fig:dofs_nedelec}
\end{figure}
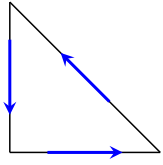

The FEM restricts the weak formulation \eqref{eq:pbcavitevar} to a finite-dimensional space denoted $V_h$ leading to the discrete variational problem : find $\Eh \in V_h$ such that 
\begin{equation}
    \int_\dom \mu_r^{-1} \rot \Eh \cdot \rot \overline{\mathbf{v}}_h - \int_\dom k^2 \epsilon_r \Eh \cdot \overline{\mathbf{v}}_h = \int_\dom \Fbf \cdot \overline{\mathbf{v}}_h,
    \label{eq:cavitypbvardiscrete}
 \end{equation}
for all $\vbf \in V_h$. A popular choice of basis functions for electromagnetic simulations are edge basis functions \cite{zbMATH03653468}, denoted $\left\{ \boldsymbol{\phi_i},~ 0 \leq i < N \right\}$. Edge elements, also called Nedelec elements, are $\Hrotdom$ conforming meaning that the finite element space $V_h$ is a subspace of $\Hrotdom$. Fig \ref{fig:dofs_nedelec} represents the degrees of freedom for edge elements of the lowest order.
\subsubsection{Assembling the linear system}
By choosing basis functions as test functions and expressing $\Eh$ in the finite element basis in \eqref{eq:cavitypbvardiscrete}, we end up with a linear system:
\begin{equation}
A \Ehvec = \Fvec,
\label{eq:linearsyst}
\end{equation}
where $A \in \mathbb{C}^{N \times N}$ and $\Fvec \in \mathbb{C}^{N}$ are defined by
\begin{align*}
    A &= \left( \int_\dom \mu_r^{-1} \rot \boldsymbol{\phi_i} \cdot \rot \boldsymbol{\phi_j} - \int_\dom k^2 \epsilon_r \boldsymbol{\phi_i} \cdot \boldsymbol{\phi_j} \right)_{1 \leq i,j \leq N}, \\
    \Fvec &= \left( \int_\dom \Fbf  \cdot \boldsymbol{\phi_i} \right)_{1 \leq i \leq N},
\end{align*}
and $\Ehvec=({E_h}_1,\dots, {E_h}_N)^T \in \mathbb{C}^N
$ contains the degrees of freedom of the discrete finite element solution, \textit{i.e.}, $\Eh= \sum_{i=1}^N {E_h}_i \boldsymbol{\phi_i}$. 

\subsubsection{Solving the linear system}
With classical computing resources, the linear system \eqref{eq:linearsyst} can be solved using iterative solvers like GMRES \cite{zbMATH01953444}. Indeed, the matrix $A$ is indefinite and possibly non-Hermitian (in the case where $\epsilon_r \notin \mathbb{R}$).

\subsection{Adaptive mesh refinement}
In the previous section, we covered how Maxwell's equations can be rewritten until finding a solution on a discretized mesh with the FEM. However, when little is known about the system at hand, the mesh is usually initialized arbitrarily. In order to have a good convergence of the solution, we need to refine the cells of the mesh specifically where it will improve the precision of the numerical scheme. In order to do so, we rely on adaptive mesh refinement strategies. More precisely, we focus on a refining strategy that uses a posteriori error estimates. We then select the cells to be refined according to a refining criterion and iterate on the mesh.
\subsubsection{A posteriori error estimation}
Let's start by looking at how we can estimate locally (in each cell) the error made compared to the exact unknown solution. The difficulty is of course that the exact solution is unknown. For problem \eqref{eq:pbcavitevar}, \cite{doi:10.1137/050636012} proposes the following residual-based error estimator $\eta$ that can be decomposed into contributions from each element $K$. Each local term is itself decomposed into 4 terms:

\begin{equation}
\eta =  \left\{ \sumK \eta_{1K}^2 + \eta_{2K}^2 + \eta_{3K}^2 + \eta_{4K}^2 \right\}^{1/2}\text.
\end{equation}

\begin{figure}[h!]
\begin{center}
\begin{tikzpicture}
\draw (0,1) -- (1.5,2) -- node[left=0.1] {$f$} (1.5,0) -- cycle;
\draw (3,1) -- (1.5,2) -- (1.5,0) -- cycle;
\draw[->] (1.5,1.25)  -- node[pos=0.75, above=0.1] {$\normal_{K_1}$} ++(0.5,0);
\draw[->] (1.5,0.75)  -- node[pos=0.75, below=0.1] {$\normal_{K_2}$} ++(-0.5,0);
\node at (0.5,1) {$K_1$};
\node at (2.5,1) {$K_2$};
\end{tikzpicture}
\end{center}
\caption{An interior face $f$ is shared by $K_1$ and $K_2$}
\label{fig:def_jump}
\end{figure}
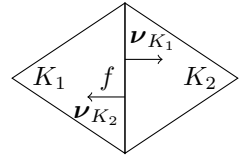

In order to define $\eta$, the following notation is introduced. Let $h_K$ denote the diameter of an element $K \in \mesh$ with outer unit normal $\normal_K$. Its boundary denoted $\partial K$ is composed of faces $f$ (or edges if $d=2$) with diameter $h_{f}$. For a face $f$ located in the interior of $\dom$ shared by two elements $K_1$ and $K_2$ (see Fig. \ref{fig:def_jump}), the jump operators are defined by

 \begin{align}
     \jumpT{\vbf} &= \vbf_{|_{K_1}} \times \normal_{K_1} +  \vbf_{|_{K_2}} \times \normal_{K_2} \\
     \jumpN{\vbf} &= \vbf_{|_{K_1}} \cdot \normal_{K_1} +  \vbf_{|_{K_2}}
      \cdot \normal_{K_2} 
 \end{align}
where the subscript $|_{K_i}$ for $i=1,2$ denotes the restriction to element $K_i$. 

The first term $\eta_{1K}^2$ can be understood as how well the propagation equation is solved:
\begin{equation}
\eta_{1K}^2 := h_K^{2} \normLtwoKvec{\Fbf - \rot \mu_r^{-1} \rot \Eh + k^2 \epsilon_r \Eh}^2.
\end{equation}
The second term $\eta_{2K}^2$ can be understood as how well the divergence of the propagation equation is solved:
\begin{equation}
\eta_{2K}^2 := h_K^2 \normLtwoK{k^2 \div(\epsilon_r \Eh)}^2. \\
\end{equation}
The third term $\eta_{3K}^2$ accounts for the tangential discontinuity of the curl of the field on each interface between the cells:
\begin{equation}
\eta_{3K}^2 :=\frac{1}{2} {\underset{f \in \partial K \backslash \froemint}{\sum}} h_{f} \normLtwovec{\jumpT{\mu_r^{-1} \rot \Eh }}
{f}^2 .\\
\end{equation}
The fourth term $\eta_{4K}^2$ takes into account the normal discontinuity on each interface between the cells:
\begin{equation}
\eta_{4K}^2 := \frac{1}{2} {\underset{f \in \partial K \backslash \froemint}{\sum}} h_{f} k^2 \normLtwovec{\jumpN{ \epsilon_r \Eh }}{f}^2.
\end{equation}

The factor $\frac{1}{2}$ in $\eta_{3K}$ and $\eta_{4K}$ is to split the contribution of an interior face between the two elements sharing this face. This allows to define a local error estimator $\eta_K$ as the sum of the four contributions $\eta_K^2 = \eta_{1K}^2 + \eta_{2K}^2 + \eta_{3K}^2 + \eta_{4K}^2$.

\subsubsection{Justification for the use of a posteriori estimates of the error}
The element estimators $\eta_K$ do not depend on the unknown exact solution $\E$ for their definition. They still estimate the error $\mathbf{e} = \E - \Eh$. Firstly, $\mathbf{e}$ is bounded from above with respect to $\eta$. This property is called reliability.

\begin{Th}[Thm 3.3 in \cite{doi:10.1137/050636012}]
There exists a constant $C$ independent of $\mesh$ such that 
\begin{equation}
    ||\mathbf{e}||^2_{\Hrotdom} \leq C \eta^2.
\end{equation}
\end{Th}

Secondly, the error is also bounded from below with respect to the a posteriori estimate. 
\begin{Th}[Thm 3.4 in \cite{doi:10.1137/050636012}]
There exists a constant $C$ independent of $\mesh$ such that 
\begin{equation}
   \eta^2 \leq ||\mathbf{e}||^2_{\Hrotdom} + \sum_{K \in \mesh} h_K^2  \normLtwovec{\Fbf - \mathcal{P}_K \Fbf}{K}^2,
\end{equation}
where $\mathcal{P}_K$ is the $L^2$ projection on $\prod_{K\in \mesh} \mathbb{P}^d_k(K)$ with $\mathbb{P}^d_k(K)$ denoting polynomials of order k.
\end{Th}
The second part in the right hand side cancels when $\Fbf$ is a polynomial of order $k$.
Since the element estimators are local, they can be used for adaptive local refinement to mark elements that should be refined. The evaluation cost on a classical computer is cheap thanks to this locality.

\subsubsection{Selection criteria}
After meshing the domain, solving the linear system and computing the error estimates, elements to be refined are marked. One possible marking strategy, the Maximum Strategy \cite[Section 7]{zbMATH05707381}, consists in selecting elements in the set
\begin{equation}
    \left\{ K \in \mesh ~|~ \eta_K \geq \theta ~ \underset{K' \in \mesh}{\max} ~\eta_{K'} \right\}
    \label{eq:refinement_criterion}
\end{equation}
where $\theta \in [0,1]$ is an input parameter. The marked elements, \textit{i.e.}, the ones with the largest error estimators, are refined. All of these steps are iterated until a criterion is reached. The procedure is illustrated in the following diagram.

\begin{center}
\begin{tikzpicture}[auto, node distance=2cm,>=latex']
    \node [name=mesh] (mesh){mesh};
    \node [right of=mesh] (solve) {solve};
    \node [right of=solve] (estimate) {estimate};
    \node [right of=estimate] (mark) {mark};
    \node [right of=mark] (refine) {refine};
    \draw [->] (mesh) -- (solve);
    \draw [->] (solve) -- (estimate);
    \draw [->] (estimate) -- (mark);
    \draw [->] (mark) -- (refine);
    \draw[->] (refine) --++(0,-2em) -| (solve);
\end{tikzpicture}
\end{center}

\section{Adaptive Mesh Refinement with Quantum Computers}
\label{sec:hybrid_amr}

Now that we have derived the classical strategy to adaptively refine meshes for the FEM, we turn towards the proposal of a numerical scheme to perform the same operations with quantum computers. We first propose a reformulation of the error estimators so that they can be sampled with quantum circuits. We then discuss how to estimate the norm of the solution depending on the quantum linear system solver.

\subsection{Reformulation of the error estimator}
\label{subsec:expression_error_estimator}
The error estimators can be reformulated as linear combinations of expectation values. For the sake of simplicity, we assume that $\epsilon_r \kappa^2 = 1$ and $\mu_r = 1$. For $\eta_{1K}^2$, we introduce the matrices $M_K$, $S_K$ and $C_K$ defined by 
\begin{align*}
{M_K} &= \left( h_K^{2} \int_K \boldsymbol{\phi_i} \cdot \boldsymbol{\phi_j} \right)_{1 \leq i,j \leq N}, \\
{S_K} &= \left( h_K^{2} \int_K \boldsymbol{\phi_i} \cdot  \rot  \rot  \boldsymbol{\phi_j} \right)_{1 \leq i,j \leq N}, \\
{C_K} &= \left( h_K^{2} \int_K \rot \rot \boldsymbol{\phi_i} \cdot \rot \rot  \boldsymbol{\phi_j} \right)_{1 \leq i,j \leq N}.
\end{align*}
By expanding the norm and using the previous matrices, we get
\begin{align*}
\eta_{1K}^2  = & h_K^{2} \left[ \int_K \Fbf \cdot \overline{\Fbf} + \int_K \rot \rot \Eh \cdot \rot \rot \overline{\E}_h \right. \\
& + \int_K \Eh \cdot \overline{\E}_h - 2 \Re \left( \int_K \Fbf \cdot \rot \rot \overline{\E}_h \right) \\
& \left. + 2 \Re \left(\int_K \Fbf \cdot \overline{\E}_h \right) - 2 \Re \left( \int_K \rot \rot \Eh \cdot \overline{\Eh} \right) \right] \\
 = & \Fvec^\dagger M_K \Fvec + \Ehvec^\dagger C_K \Ehvec + \Ehvec^\dagger M_K \Ehvec  \\
& - 2 \Re(\Fvec ^\dagger S_K \Ehvec) + 2 \Re(\Fvec^\dagger M_K \Ehvec) \\
&  - 2 \Re(\Ehvec^\dagger S_K \Ehvec) .
\end{align*}

Similarly, $\eta_{2K}$ can be formulated as
\begin{equation*}
\eta_{2K}^2 = \Ehvec^\dagger D_K \Ehvec, \\
\end{equation*}
where 
\begin{equation*}
{D_K} = \left( h_K^{2} \int_K \div \boldsymbol{\phi_i} \div \boldsymbol{\phi_j} \right)_{1 \leq i,j \leq N}. \\
\end{equation*}
For $\eta_{3K}$, contributions of the neighboring elements appear because of the jump. By definition of the jump, it holds that
\begin{equation*}
\eta_{3K}^2 = \frac{1}{2} \underset{K \cap K'=f}{\sum_{f\in\partial K \backslash \froemint}} h_{f} \normLtwovec{ ( \rot \Eh \times \normal)_{|_{K}} + ( \rot \Eh \times \normal)_{|_{K'}} }{f}^2.
\end{equation*}
For two elements $K_1$ and $K_2$ sharing a face $f$, let $C_{K_1 K_2 f}$ be the matrix defined by
\begin{equation*}
{C_{K_1 K_2 f}} = \left( h_{f} \int_f ( \rot \boldsymbol{\phi_i} \times \normal)_{|_{K_1}} \cdot ( \rot \boldsymbol{\phi_j} \times \normal)_{|_{K_2}} \right)_{1 \leq i,j \leq N}. \\
\end{equation*}
By expanding the norm, $\eta_{3K}$ can be rewritten as
\begin{align*}
\eta_{3K}^2 = \frac{1}{2} \underset{K \cap K'=f}{\sum_{f\in\partial K \backslash \froemint}} & \left[ \Ehvec^\dagger C_{KKf} \Ehvec + \Ehvec^\dagger C_{K'K'f} \Ehvec \right. \\
& \left. + 2\Re(\Ehvec^\dagger C_{KK'f} \Ehvec) \right].
\end{align*}
The process is similar for $\eta_{4K}$ leading to
\begin{align*}
\eta_{4K}^2 = \frac{1}{2} \underset{K \cap K'=f}{\sum_{f\in\partial K \backslash \froemint}} & \left[ \Ehvec^\dagger M_{KKf} \Ehvec + \Ehvec^\dagger M_{K'K'f} \Ehvec \right. \\
& \left. + 2\Re(\Ehvec^\dagger M_{KK'f} \Ehvec)  \right],
\end{align*}
where
\begin{align*}
{M_{K_1 K_2 f}} &= \left( h_{f} \int_f ( \boldsymbol{\phi_i} \cdot \normal)_{|_{K_1}} ( \boldsymbol{\phi_j} \cdot \normal)_{|_{K_2}} \right)_{1 \leq i,j \leq N}.
\end{align*}
Combining these four quantities, one element estimator $\eta_K$ contains 31 terms for $d=3$ and 25 terms for $d=2$.

\subsection{Quantum circuits for error estimation}
\label{subsec:strategy}
The terms presented in section \ref{subsec:expression_error_estimator} can be separated into three categories $ \Ehvec M \Ehvec$ (bilinear), $\Fvec^\dagger M \Fvec$ (constant) and $\Ehvec^\dagger M \Fvec$ (linear). Several points need particular focus to be able to estimate these terms. Quantum computing relies on normalized states and unitary operators. However the vectors $\Ehvec$ and $\Fvec$ do not have euclidian norm equal to one and the matrices are not unitary. \\
We rely on block-encoding, a technique that allows to embed a subnormalized matrix as the top-left block of a unitary \cite{Gilyen_2019}. 

\begin{figure}[h!]
\centering
\includegraphics[scale=0.4]{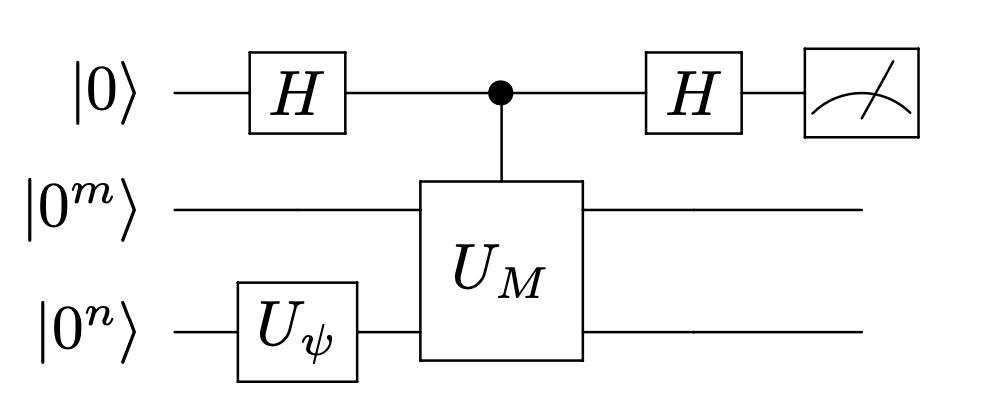}
\caption{Circuit for a Hadamard test to estimate $\frac{1}{\alpha} (\Re(\bra{\psi} M \ket{\psi})$. The unitary $U_{\psi}$ is defined by $U_{\psi} \ket{0} = \ket{\psi}$.}
\label{fig:Hadamard_diag}
\end{figure}

\begin{figure}[h!]
\centering
\includegraphics[scale=0.35]{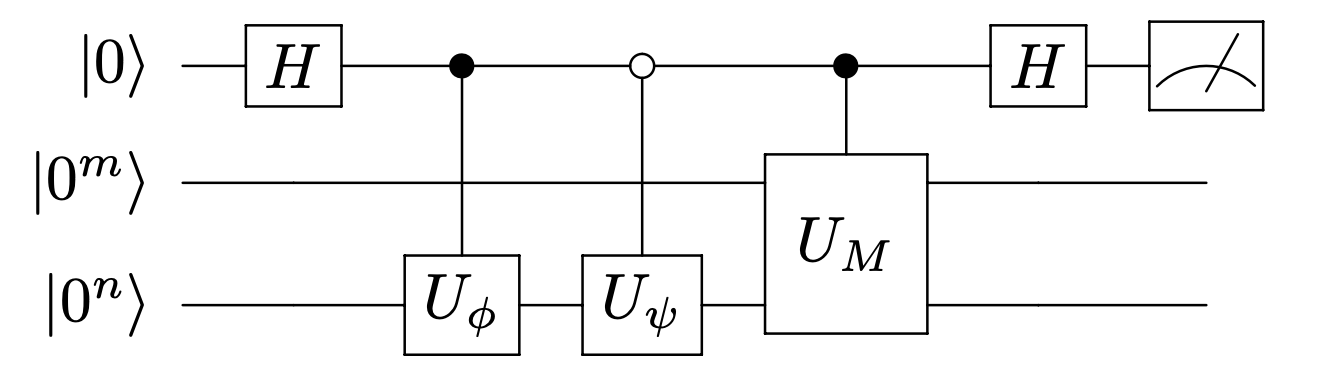}
\caption{Circuit for a Hadamard test to estimate $\frac{1}{\alpha} \Re(\bra{\phi}M\ket{\psi})$. The unitaries $U_{\phi}$ and $U_{\psi}$ prepare respectively the states $\ket{\phi}$ and $\ket{\psi}$.}
\label{fig:Hadamard_off_diag}
\end{figure}

\begin{Def}[Block-encoding \cite{Gilyen_2019, lin2022lecturenotesquantumalgorithms}]
Given an n-qubit matrix $M$, an $(\alpha,m,\epsilon)$-block-encoding of $M$ is an (m+n)-qubit unitary matrix $U_M$ such that
\begin{equation*}
||M-\alpha(\bra{0^m} \otimes I_n)U_M(\ket{0^m}\otimes I_n)|| \leq \epsilon.
\end{equation*}
\end{Def}
The parameter $\alpha$ is the subnormalization factor, $m$ is the number of ancilla qubits in the block-encoding and $\epsilon$ is the precision. Supposing access to block-encodings of the matrices $M_K$, $S_K$, $C_K$, $D_K$, $C_{KK'f}$ and $M_{KK'f}$, Hadamard tests can be used to estimate the different expectation values \cite{PhysRevA.104.032422}.
We suppose access to a quantum linear system solver that outputs $\ket{E_h}$ and to a unitary $U_F$ preparing $\ket{F}$.
Given $U_M$ a $(\alpha,m,\epsilon)$-block-encoding of $M$, the circuit from Fig. \ref{fig:Hadamard_diag} allows to estimate $\frac{1}{\alpha} \Re(\bra{E_h}M \ket{E_h})$ or  $\frac{1}{\alpha} \Re(\bra{F}M\ket{F}))$. The circuit from Fig. \ref{fig:Hadamard_off_diag} can be used to estimate $\frac{1}{\alpha} \Re(\bra{E_h}M\ket{F})$. More precisely, if $U_M$ is a $(\alpha,m,0)$-block-encoding, the probabilities of observing 0 when measuring the ancilla qubit are $\frac{1}{2} \left( 1 + \frac{1}{\alpha}\Re(\bra{\psi} M \ket{\psi}) \right)$ and $\frac{1}{2} \left( 1 + \frac{1}{\alpha}\Re(\bra{\phi} M \ket{\psi}) \right)$.\\

To approximately compute the error estimators using the results from the Hadamard tests, the euclidian norms of $\Fvec$ and $\Ehvec$ as well as the subnormalization factors of the different block-encodings are required. We can assume access to $||\Fvec||$ since $\Fbf$ is an input of the problem. The norm of $\Ehvec$ can be approximately computed. The procedure, which depends on the quantum solver, is detailed in the  next section. The values of $\alpha$ depend on the norms of the block-encoded matrices and on the particular construction of the block-encodings.

\subsection{Quantum linear system solver}
\label{subsec:quantum_linear_solver}
Both non-variational and variational quantum linear system solvers can be used to approximate $\ket{E_h}$. But the procedure to estimate the norm depends on the solver.
\subsubsection{VQLS}
The variational quantum linear solver (VQLS) \cite{BravoPrieto2023variationalquantum} is a hybrid algorithm. The inputs of VQLS are a circuit $U_F$ preparing $\ket{F}$ and the matrix $A$ given as a linear combination of unitaries. A quantum parametrized circuit, called ansatz, is used to approximate the solution $\ket{E_h}$.
We choose the local cost function $C_L$ defined by
\begin{equation}
C_L= \frac{\bra{E_h} H_L \ket{E_h}}{||A\ket{E_h}||},
\end{equation}
associated with the local Hamiltonian
\begin{equation}
    H_L = A^\dagger U_F \left( \mathbb{1} - \frac{1}{n} \sum_{j=1}^{n} \ket{0_j}\bra{0_j} \otimes \mathbb{1}_{\bar{j}} \right) U_F^\dagger A,
\end{equation}
where $\ket{0_j}$ is the zero state on qubit j and $\mathbb{1}_{\bar{j}}$ is the identity applied to all qubits but j. The parameters are optimized using a classical computer. The euclidian norm of the solution $||\Ehvec||$ satisfies \cite{arora2025implementationfiniteelementmethod}
\begin{equation}
    ||\Ehvec||=\frac{||\Fvec||}{\bra{F}A\ket{E_h}},
\end{equation}
in which the quantity $\bra{F}A\ket{E_h}$ can be computed in a way similar to the cost functions in \cite{BravoPrieto2023variationalquantum}.

\subsubsection{Non-variational quantum linear system solver}
The work \cite[Thm 33]{https://doi.org/10.4230/lipics.icalp.2019.33} proposes an algorithm that outputs a state $\epsilon$-close to the solution $A^{-1}\ket{F}/||A^{-1}\ket{F}||$ with a query complexity to a $(\alpha,a,\delta)$-block-encoding of $A$ of $\bigO(\kappa \text{polylog}(\kappa/\epsilon))$ (with $\kappa$ the condition number of $A$). It is also possible to estimate the norm of the solution $||A^{-1} \ket{F}||$ with a complexity $\bigO(\kappa \epsilon^{-1} \text{polylog}(\kappa/\epsilon)))$ using variable-time amplitude estimation.

\subsection{Hybrid adaptive mesh refinement loop}
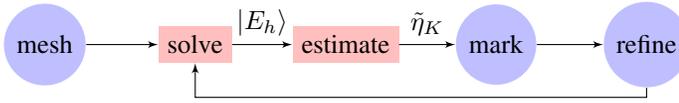
\begin{figure}[htbp]
\begin{center}
\begin{tikzpicture}[auto, node distance=2cm,>=latex',every text node part/.style={align=center}]
    \node [circle, name=mesh, fill=blue!25] (mesh){mesh};
    \node [right of=mesh, fill=red!25] (solve) {solve};
    \node [right of=solve, fill=red!25] (estimate) {estimate};
    \node [circle, right of=estimate, fill=blue!25] (mark) {mark};
    \node [circle, right of=mark, fill=blue!25] (refine) {refine};
    \draw [->] (mesh) -- (solve);
    \draw [->] (solve) -- (estimate) node[midway,above] {$\ket{E_h}$};
    \draw [->] (estimate) -- (mark) node[midway,above] {$\Tilde{\eta}_K$};
    \draw [->] (mark) -- (refine);
    \draw[->] (refine) --++(0,-2em) -| (solve);
\end{tikzpicture}
\end{center}
\caption{Hybrid adaptive mesh refinement algorithm. Blue circles symbolise steps performed on a classical computer. Red rectangles refer to steps performed on a quantum computer.}
\label{fig:hybrid_algo}
\end{figure}

A high-level overview of the hybrid adaptive mesh refinement is shown in Fig \ref{fig:hybrid_algo}. First, the step ``mesh'' consists in creating the initial mesh on a classical computer. In ``solve", the linear system \eqref{eq:linearsyst} is solved using a quantum linear system solver. The quantum step ``estimate" outputs classical local error estimators $\Tilde{\eta}_{K}$ approximately computed following the quantum procedure presented in section \ref{subsec:strategy}. These are used to mark elements using the criterion \eqref{eq:refinement_criterion} in ``mark". ``Refine" splits the tagged elements on a classical computer.

\subsection{Query complexity to a QLSP solver}
For a mesh made up of $N_{\text{elements}}$ elements, there are $N_{\text{elements}}$ local error estimators to approximate.
For each one of them, to reach precision $\epsilon$ on the squared local estimator, i.e. $|\eta_K^2 - \Tilde{\eta}_K^2| \leq \epsilon$, one needs a precision $\frac{\epsilon}{\max \alpha ||\Ehvec||^2}$ on the results of the Hadamard tests. As a consequence $\bigO(\frac{(\max \alpha)^2 ||\Ehvec||^4}{\epsilon^2})$ samples are required. The query complexity to the quantum linear system solver is $\bigO( \frac{N_{\text{elements}} (\max \alpha)^2 ||\Ehvec||^4}{\epsilon^2})$. For edge elements of lowest order, $\bigO
(N_{\text{elements}}) = \bigO(N)$ (with $N$ the size of the problem) and the query complexity is $\bigO( \frac{N (\max \alpha)^2 ||\Ehvec||^4 }{\epsilon^2})$. 

\section{Numerical simulations}
\label{sec:numerical_simulations}
We now turn to numerical experiments that support our theoretical findings.
In order to be able to run experiments with reduced emulation resources, we choose to test VQLS as the linear system solver.

\subsection{Software}
The Python library DOLFINx \cite{BarattaEtal2023} is used for the FEM simulations. The procedure detailed in \ref{subsec:strategy} is implemented using the Qiskit framework \cite{qiskit2024}. The different block-encodings are obtained using the FABLE framework \cite{9951292}. 
VQLS simulations are ran using vqls-prototype~\cite{vqls_prototype}.

\subsection{Error estimators}
\begin{figure}[h!]
    \centering
    \begin{subfigure}{0.23\textwidth}
    \includegraphics[scale=0.064]{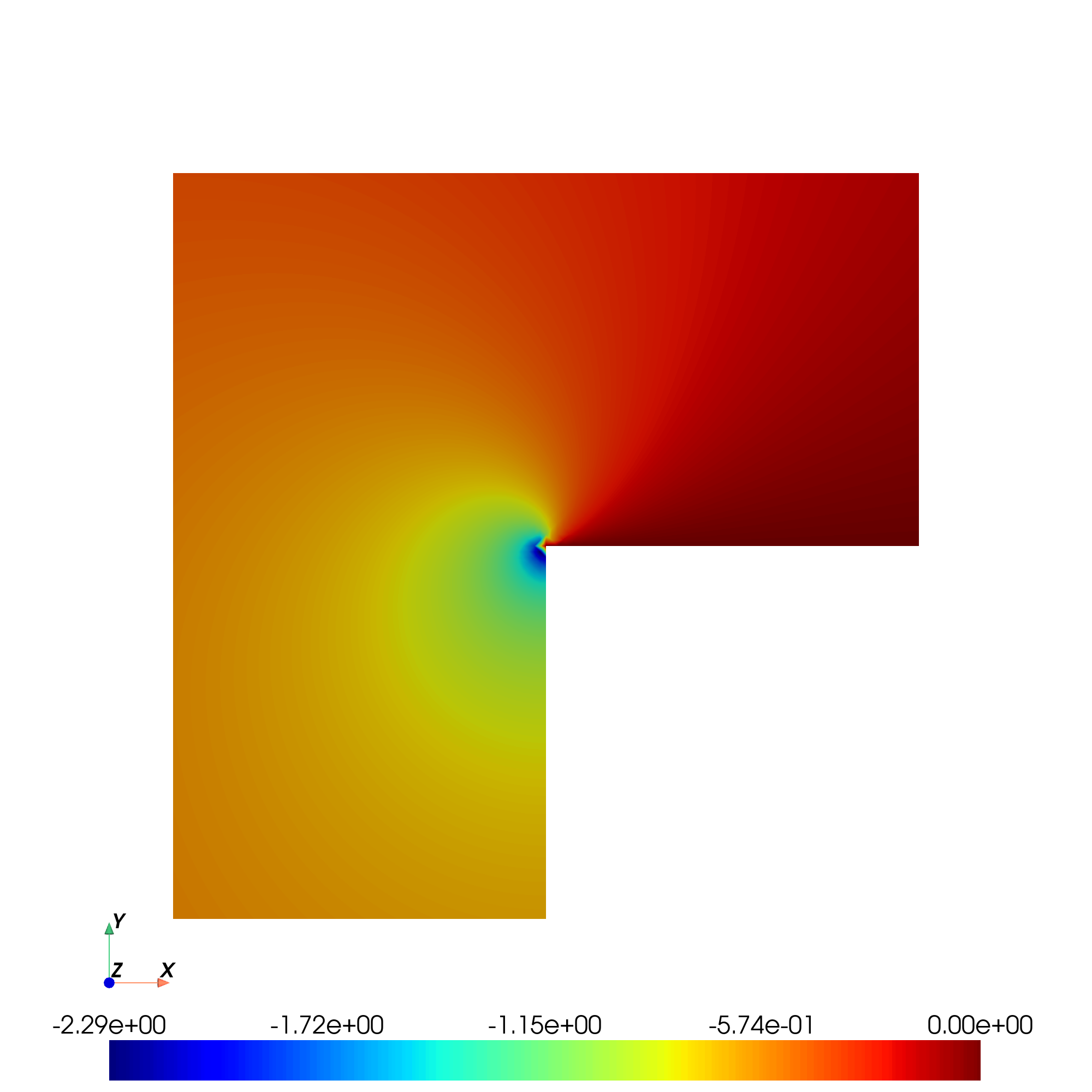}
    \end{subfigure}
    \begin{subfigure}{0.23\textwidth}
    \includegraphics[scale=0.064]{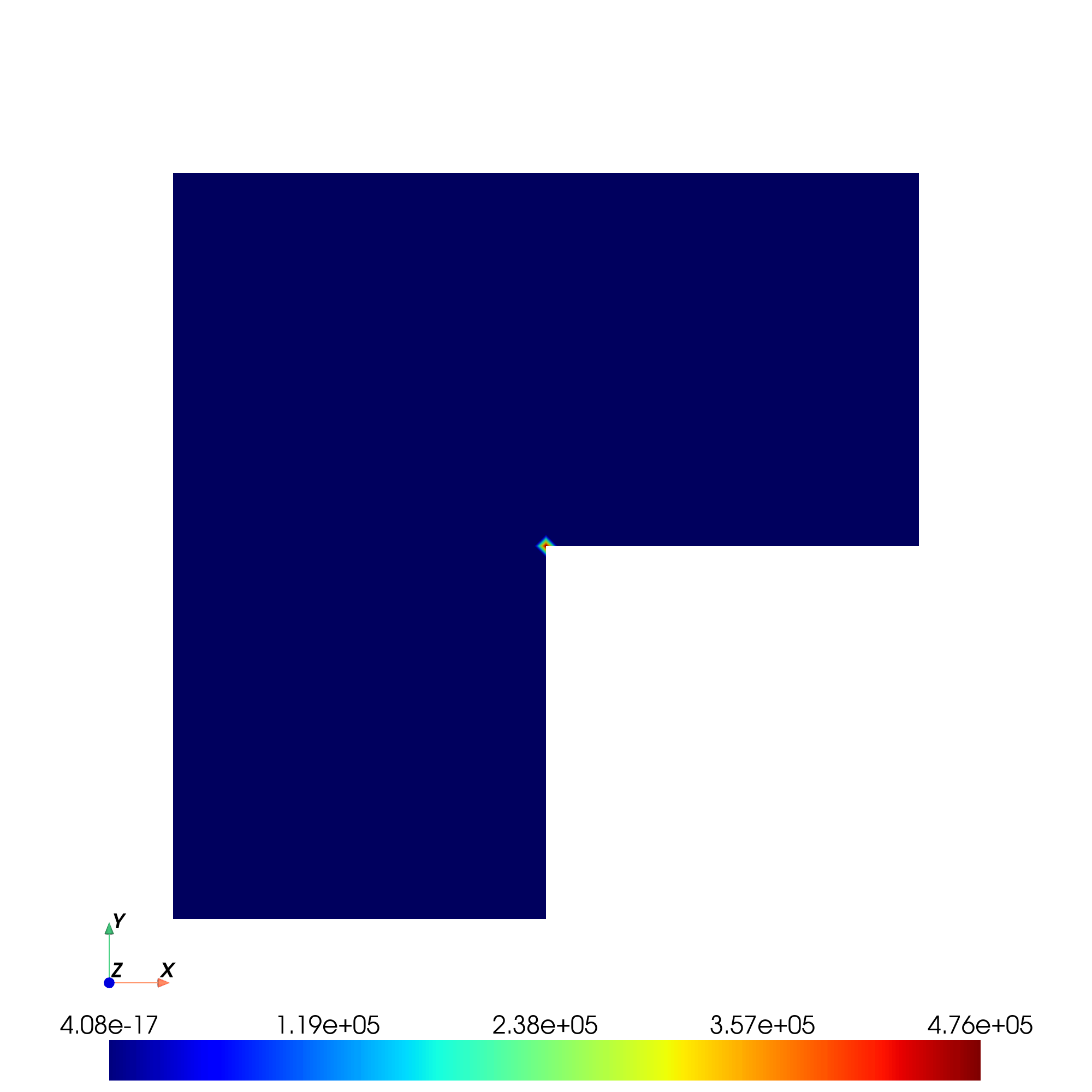}
    \end{subfigure}
    \caption{Components along the x-axis (left) and y-axis (right) of the exact solution. The exact solution is interpolated on the finest uniform mesh using order one Lagrange finite elements.}
    \label{fig:exact_solution}
\end{figure}

Simulations are performed in a two-dimensional L-shaped domain $\dom = [-1,1]^2\backslash [0,1] \times [-1,0]$. We choose the right hand side and the boundary condition such that $\E = \grad(r^{2/3}\sin(2\phi/3))$ (in polar coordinates) is the exact solution. This solution, which is depicted in Fig \ref{fig:exact_solution}, diverges at the reentrant corner where $r=0$.

The discretization uses edge elements of lowest order. We choose $\theta=0.6$ for the selection criterion \eqref{eq:refinement_criterion} and the algorithm stops when ten adaptive iterations have been performed. There are 22 degrees of freedom on the input mesh meaning that $n=5$ qubits encode the solution. Because $2^5=32$, 22 amplitudes encode the degrees of freedom, the rest of the amplitudes are padded with zeros.

\begin{figure}[h!]
    \centering
    \begin{subfigure}{0.23\textwidth}
    \includegraphics[scale=0.064]{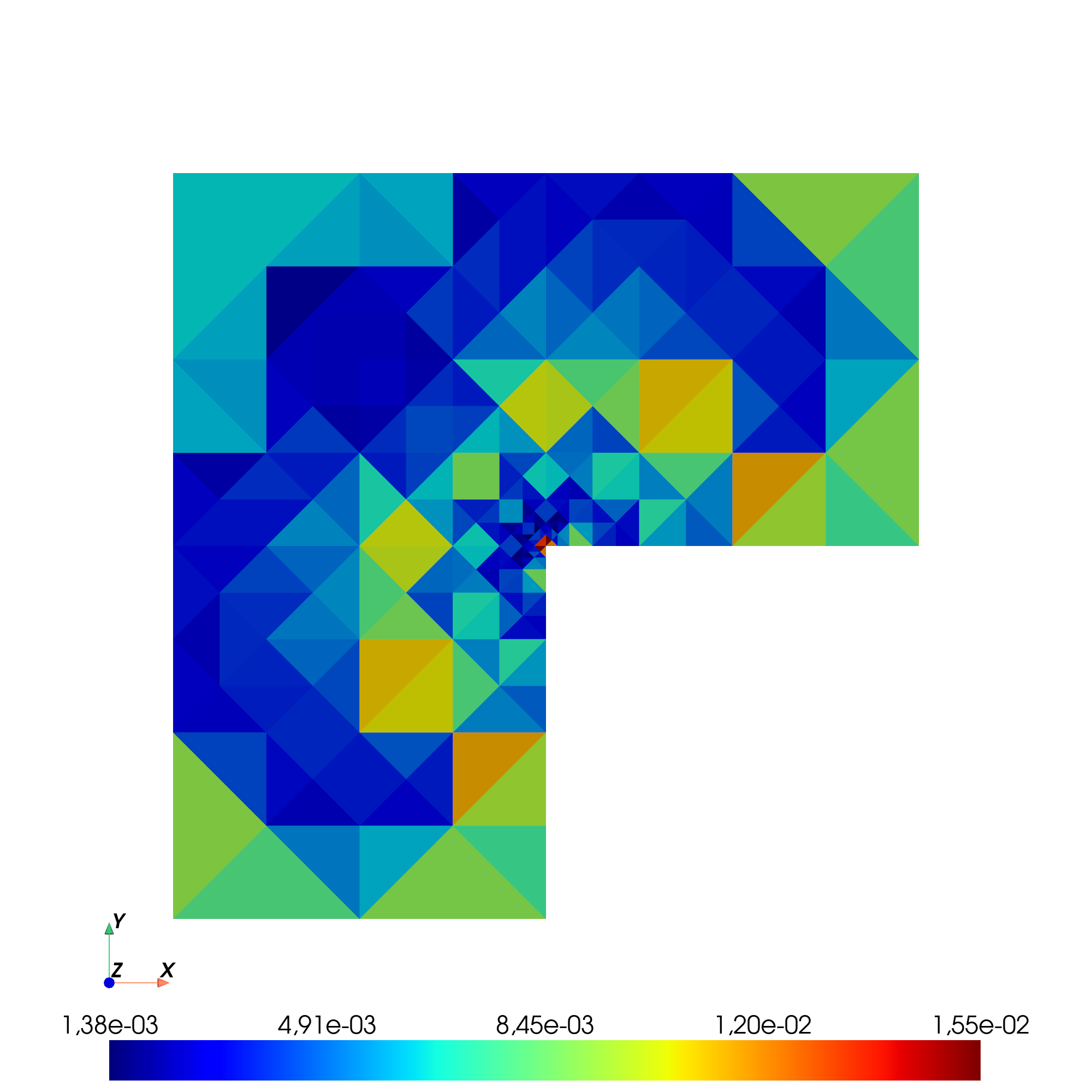}
    \end{subfigure}
    \begin{subfigure}{0.23\textwidth}
    \includegraphics[scale=0.064]{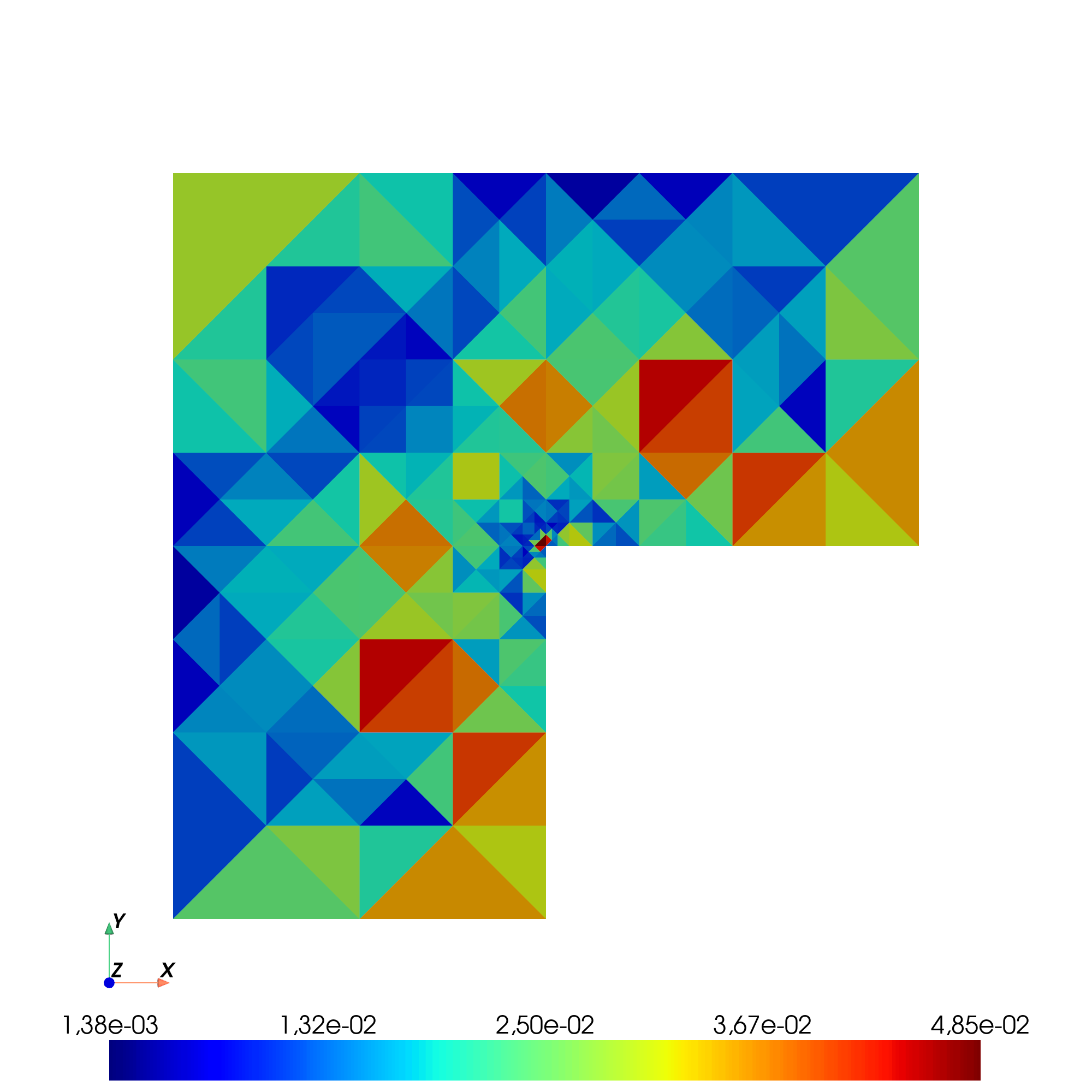}
    \end{subfigure}
    \caption{Local $\Hrot$ error (left) and local error estimator $\eta_K$ (right) at iteration $5$ in the classical refinement loop.}
    \label{fig:error_estimator_level_5}
\end{figure}

\begin{figure}[h!]
    \centering
    \begin{subfigure}{0.23\textwidth}
    \includegraphics[scale=0.065]{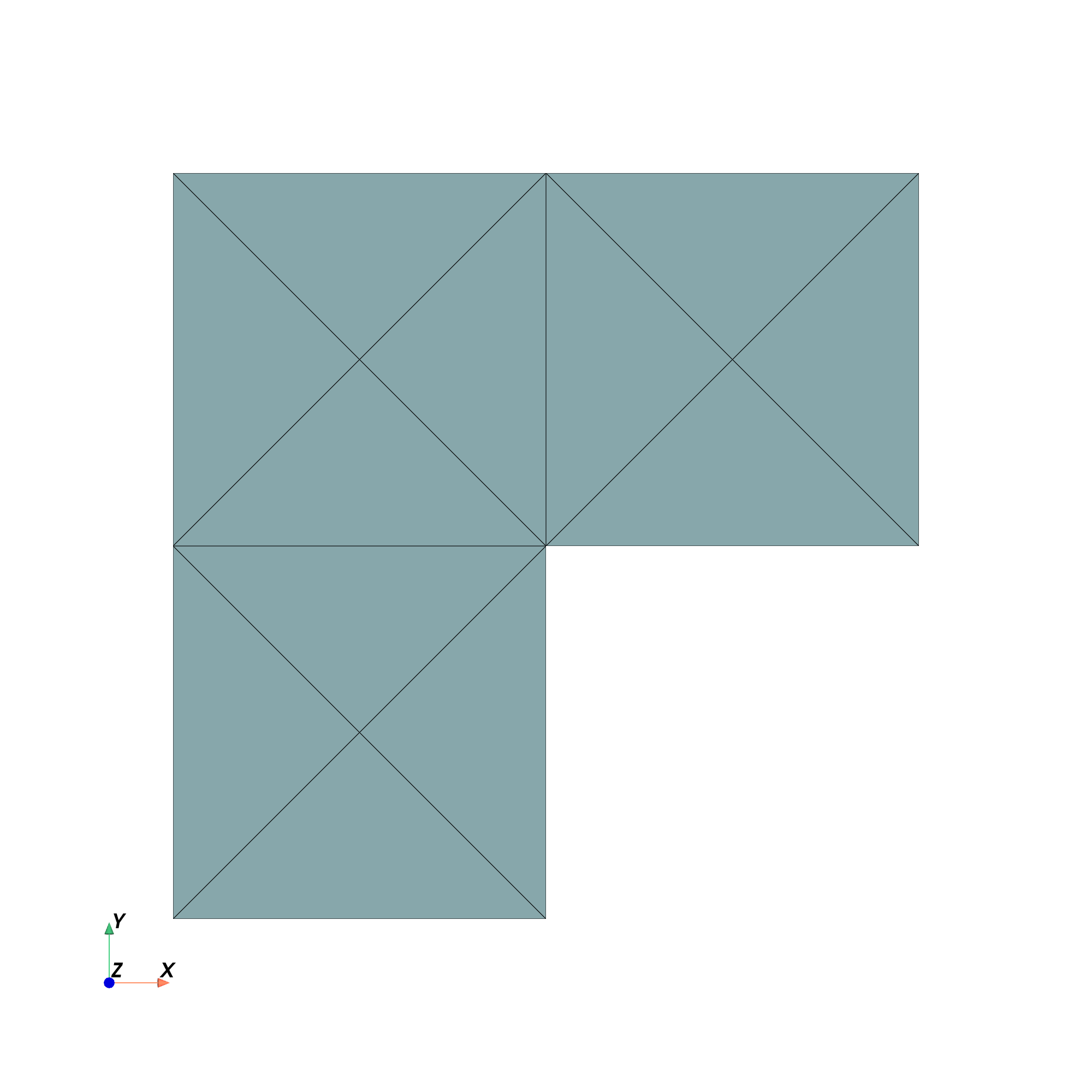}
    \end{subfigure}
    \begin{subfigure}{0.23\textwidth}
    \includegraphics[scale=0.065]{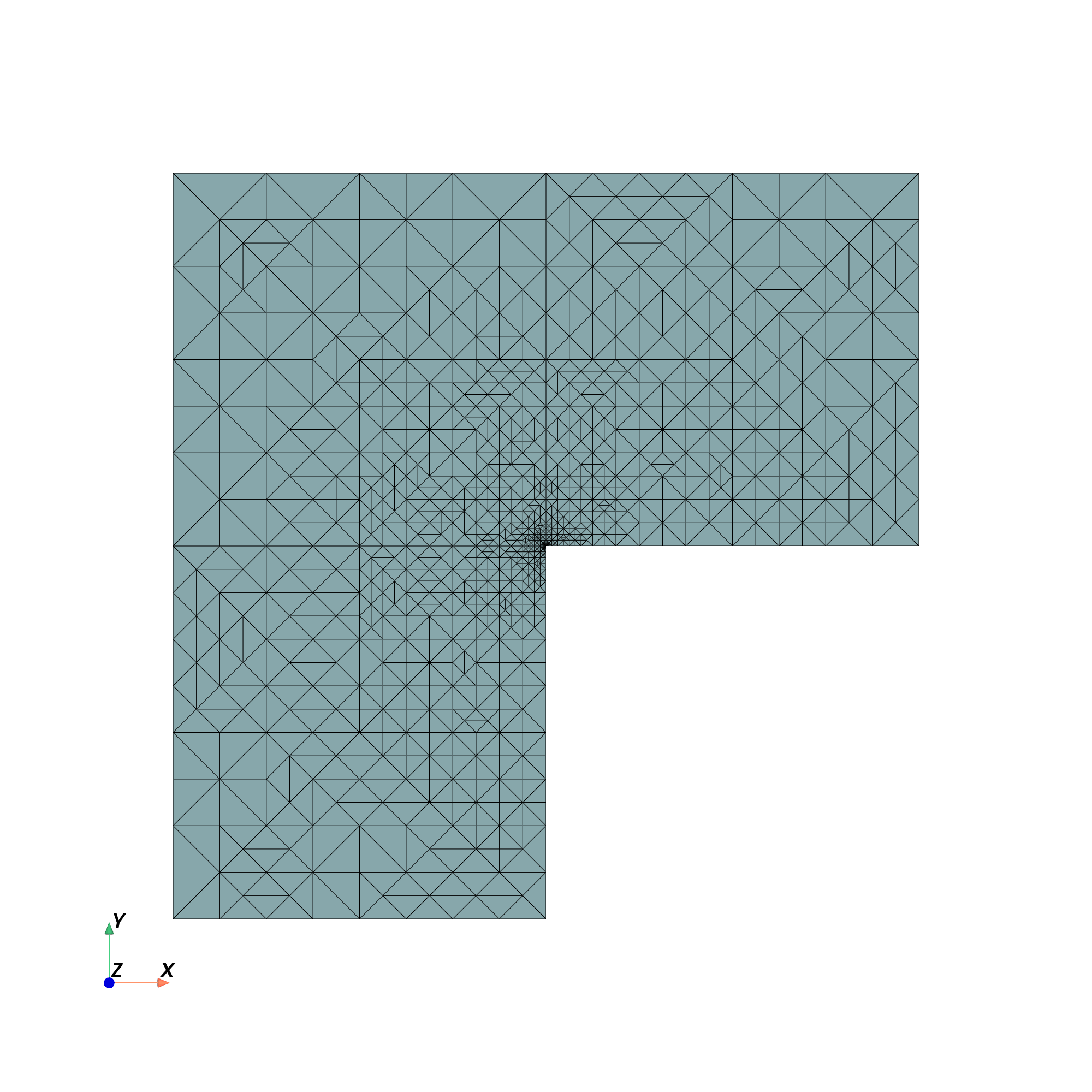}
    \end{subfigure}
    \caption{Initial mesh (left) and mesh at the final iteration (right).}
    \label{fig:initial_mesh_final_mesh_classical}
\end{figure}

The local error estimators $\eta_K$ are good approximations for the usually unknown local $\Hrot$ errors 
\begin{equation}
|| \mathbf{e} ||_{\Hrotspace{K}} = \left( \normLtwoKvec{\mathbf{e}}^2 + \normLtwoKvec{\rot \mathbf{e}}^2 \right)^{1/2}
\end{equation}
with $\mathbf{e}=\E-\Eh$ as can be seen in Fig. \ref{fig:error_estimator_level_5}.
Both these quantities are large near the re-entrant corner where the solution is singular. This explains the concentration of elements in this region in the final mesh compared to the inital mesh (see Fig. \ref{fig:initial_mesh_final_mesh_classical}).

\begin{figure}[h!]
    \centering
    \includegraphics[scale=0.6]{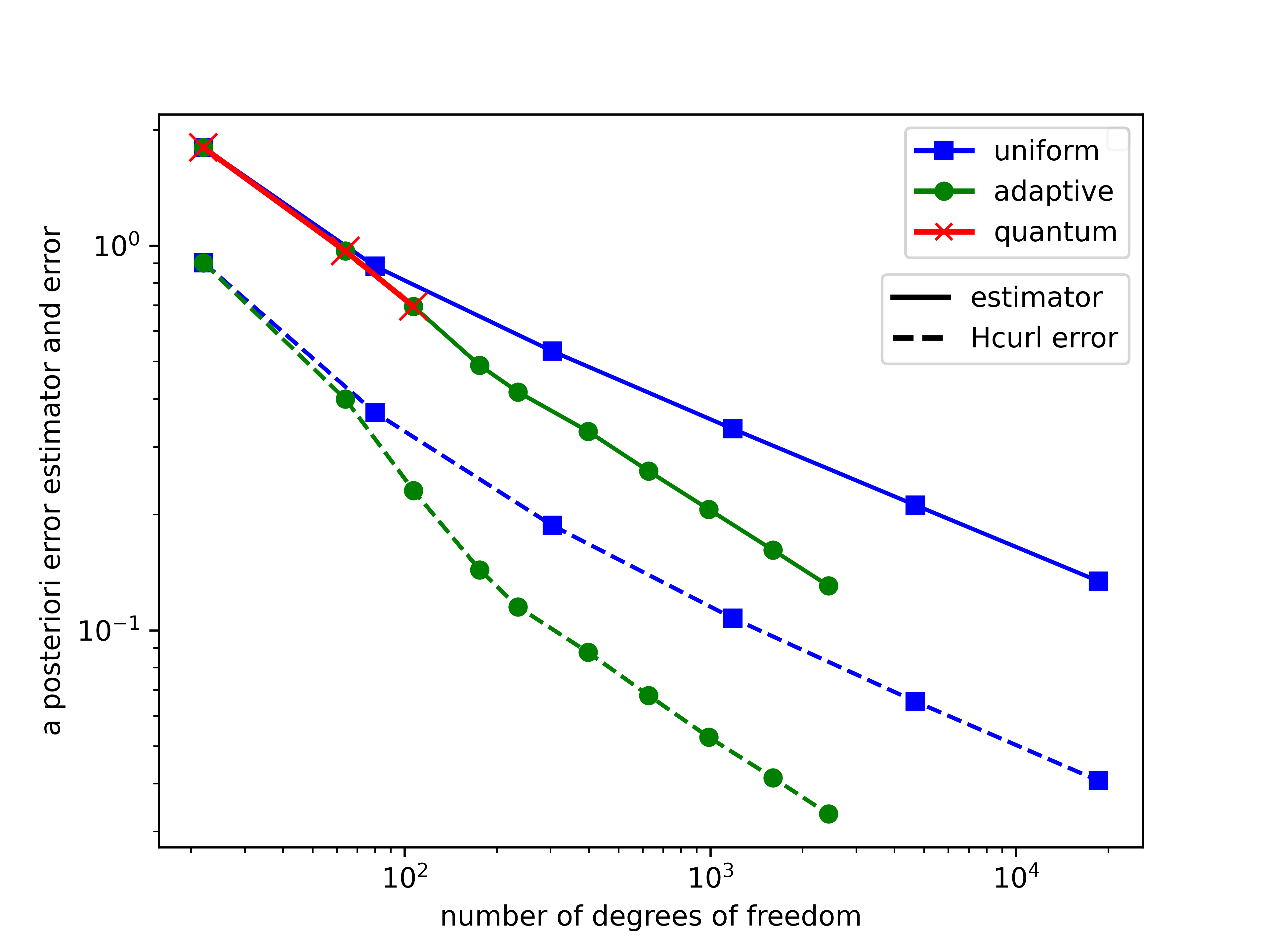}
    \caption{$\Hrotdom$ error (dashed lines) and a posteriori error estimator $\eta$ (solid lines) as a function of the number of degrees of freedom. The blue squares correspond to classical uniform mesh refinement. The green dots correspond to classical adaptive mesh refinement. The red crosses are obtained from the quantum circuits presented in section \ref{subsec:strategy}. Without sampling error, the classical and quantum error estimators are overlaid.}
    \label{fig:cv_rate_unif_adapt_classical_adapt_quantum}
\end{figure}

Fig. \ref{fig:cv_rate_unif_adapt_classical_adapt_quantum} represents the $\Hrot$ error and the a posteriori error estimator $\eta$ as a function of the number of degrees of freedom. The classical convergence rate is improved using adaptive mesh refinement rather than uniform mesh refinement (i.e. all the elements are refined). So this example illustrates the importance of classical adaptive mesh refinement. 
We are able to compute the local estimators with the quantum circuit for three iterations by encoding the classical finite element solution in a quantum state at each iteration. Without sampling error, the quantum local estimators are equal to the classical ones. This validates the correctness of the circuits. In preliminary runs taking into account the sampling error, this error dominates the estimators to the point that the result is not a valid approximation of the error.

\subsection{Experiments with VQLS}
\begin{figure}[h!]
\centering
\begin{subfigure}{0.22\textwidth}
\includegraphics[scale=0.28]{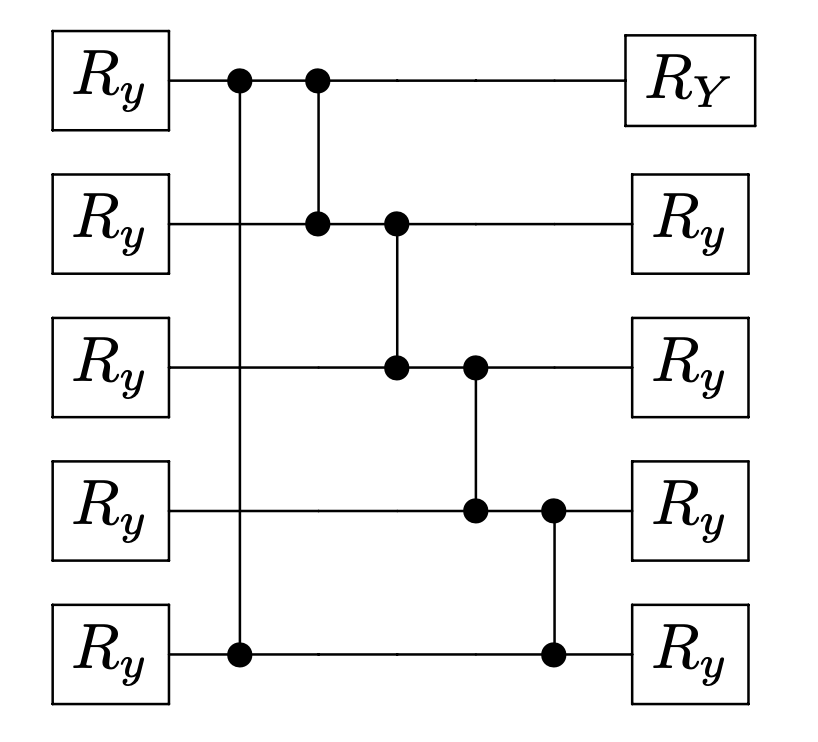}
\end{subfigure}
\begin{subfigure}{0.22\textwidth}
\includegraphics[scale=0.28]{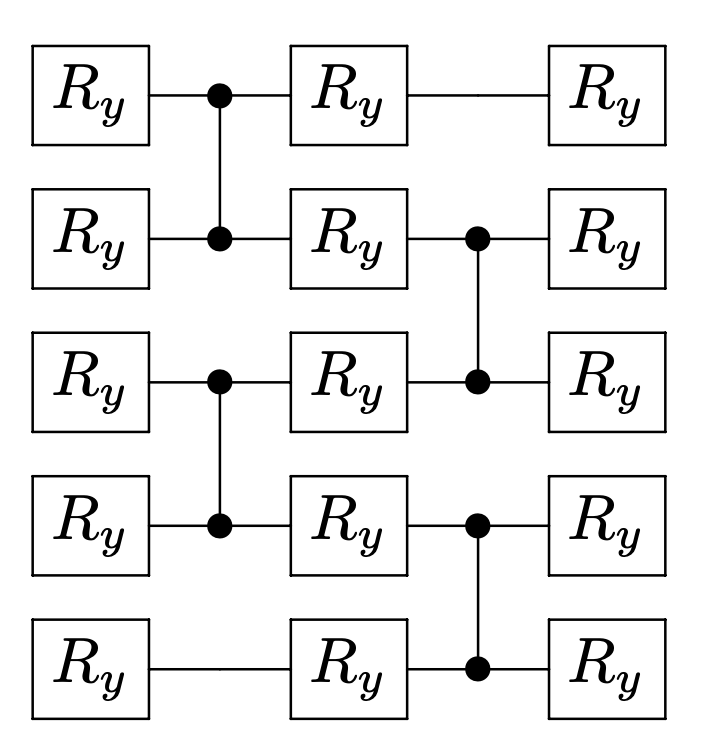}
\end{subfigure}
\caption{Linear circular $R_y$-$CZ$ ansatz \cite{Sim_2019} (left) and linear alternating $R_y$-$CZ$ ansatz \cite{BravoPrieto2023variationalquantum} (right) represented for $n=5$ qubits and one layer.}
\label{fig:ansatzes}
\end{figure}

In order to implement the entire workflow we need to solve the quantum linear system problem focusing on VQLS. Since the solution is real-valued, we are interested in an ansatz that produces a quantum state with real amplitudes. The two studied fixed-structure layered hardware-efficient ansatzes are represented in Fig. \ref{fig:ansatzes}.

We use the constrained optimization by linear approximation (COBYLA) algorithm \cite{zbMATH00653035} to optimize the parameters.
We denote by $\ket{{E_h}_{\text{VQLS}}}$ the VQLS output and define 
\begin{equation}
    \ket{{E_h}_{\text{FEM}}} = \sum_{i=1}^N \frac{{E_h}_i}{|| \Ehvec||} \ket{i}
\end{equation}
in order to compare both quantities.
We investigate the evolution of the fidelity $|\bra{{E_h}_{\text{VQLS}}}\ket{{E_h}_{\text{FEM}}}|^2$ with respect to the number of qubits.
To vary the problem size, we extract submatrices and subvectors from the initial 5-qubit linear system.
The number of layers in the ansatzes and the maximum number of iterations for the optimizer vary with the number of qubits.
Table \ref{tab:param_vqls} details the parameters used for the simulations.

\begin{figure}[h!]
    \centering
    \includegraphics[scale=0.6]{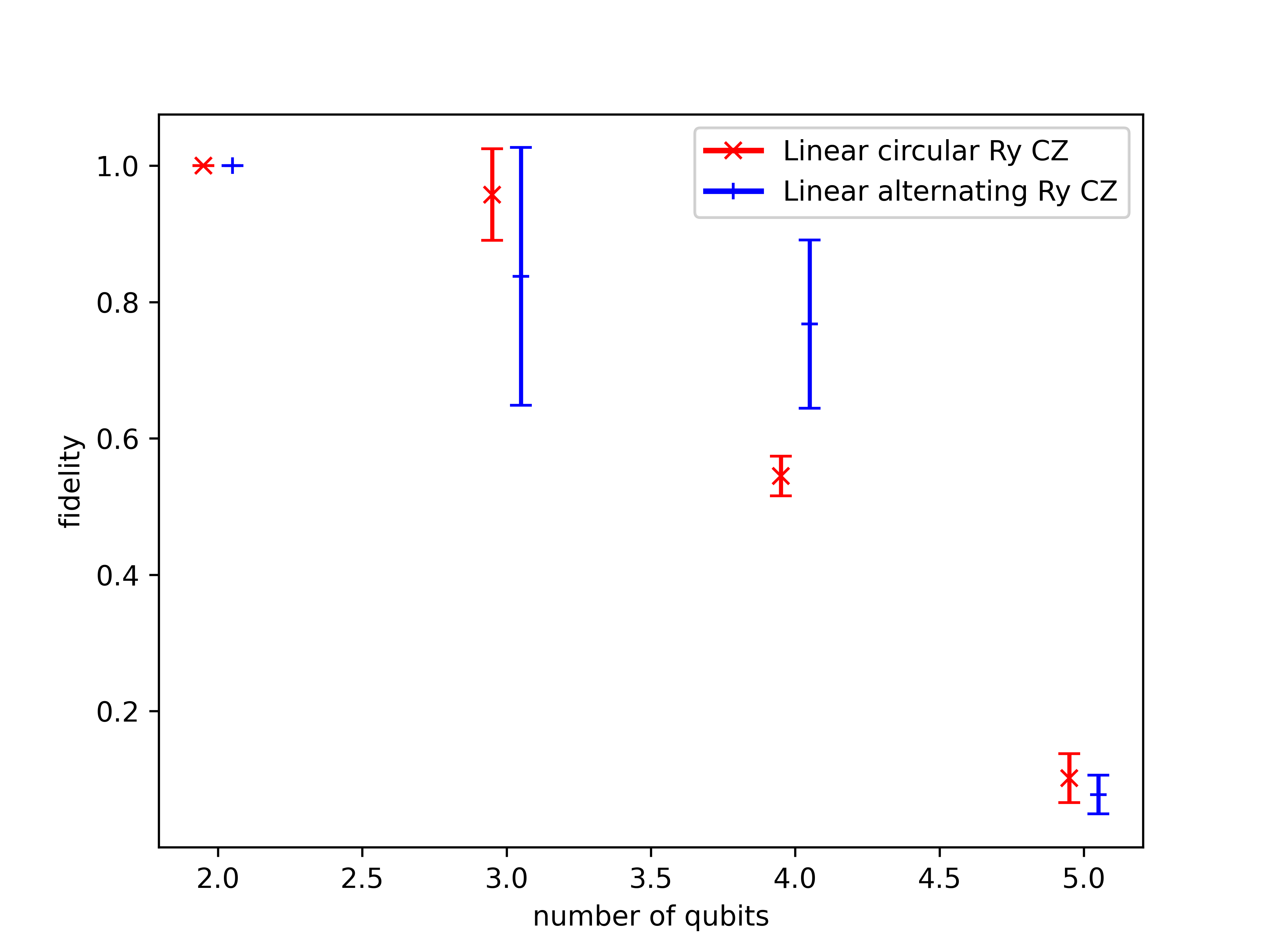}
    \caption{Mean fidelity $|\bra{{E_h}_{\text{VQLS}}}\ket{{E_h}_{\text{FEM}}}|^2$ computed with 5 simulations for a number of qubits ranging from 2 to 5 for the linear circular $Ry-CZ$ ansatz (red x) and the linear alternating $Ry-CZ$ ansatz (blue +). The 95\% confidence intervals are plotted.}
    \label{fig:fidelity_nqbits}
\end{figure}

\begin{table}[h!]
\caption{Parameters for the ansatz and the optimizer used in VQLS simulations. }
\begin{center}
\begin{tabular}{|c|c|c|c|c|}
\hline
number of qubits $n$ & 2 & 3 & 4 & 5 \\
\hline
number of layers & 1 & 2 & 4 & 8 \\
\hline
maximum number of iterations & 1000 & 2000 & 5000 & 10000  \\
\hline
\end{tabular}
\label{tab:param_vqls}
\end{center}
\end{table}
The results, without sampling error, are shown in Fig.~\ref{fig:fidelity_nqbits}.
The fidelity is computed using five simulations. The fidelity is high for two qubits. For both ansatzes, the fidelity decreases when the number of qubits increases and reaches very low values, which shows that VQLS does not perform well on our problem.

\section{Conclusion}
\label{sec:conclusion}
In this work, we have proposed an extension of an adaptive mesh refinement algorithm for time-harmonic Maxwell's equation to the quantum formalism.
In order to do so, we have first derived the propagation equation from Maxwell's equations.
Then, we covered the classical mesh refinement loop and studied its formulation in order to adapt it to the quantum formalism.
The local error estimators can be extracted as expectation values of quantum circuits.
Using block-encoding, localized a posteriori error estimators are approximately computed on quantum computers.
Our method has been validated on simulations which do not account for sampling error thus recovering the convergence rate of the classical setup. 

Until now, in the hybrid mesh adaptation loop, the quantum sampling error dominates the error coming from the discretization onto the mesh.
Future work should aim at improving this and proposing an algorithm that is efficient overall. Preconditioning of the linear system as proposed by \cite{deiml2024quantum} is a promising lead. 

\bibliographystyle{unsrt}

\bibliography{main}

\end{document}